\documentclass{article}[10pt]
\usepackage{spconf,amsmath,graphicx,epsfig,multirow, bbm, algpseudocode, algorithm, mathtools}

\DeclareMathOperator*{\var}{var}

\DeclarePairedDelimiter\floor{\lfloor}{\rfloor}

\title{Non-Adaptive Policies for 20 Questions Target Localization}

\name{Ehsan Variani$^1$, Kamel Lahouel$^2$, Avner Bar-Hen$^3$, Bruno Jedynak$^2$}
\address{$^1$ Electrical \& Computer Engineering, Johns Hopkins Univ., Baltimore, MD, USA\\
  $^2$Applied Mathematics \& Statistics, Johns Hopkins Univ., Baltimore, MD USA\\
  $^3$Université Paris Descartes, MAP5, Paris, Frane\\
  {\small \tt \{variani, klahoue1, bruno.jedynak\}@jhu.edu \quad avner.bar-hen@mi.parisdescartes.fr}}

\vspace{3.0cm}
\ninept
\begin{document}

\maketitle
\newtheorem{proposition}{Proposition}
\newtheorem{theorem}{Theorem}
\newtheorem{definition}{Definition}
\newtheorem{lemma}{Lemma}
\newtheorem{corollary}{Corollary}

\begin{abstract}

The problem of target localization with noise is addressed. The
target is a sample from a continuous random variable with known distribution
and the goal is to locate it with minimum mean squared error distortion. The
localization scheme or policy proceeds by queries, or questions, weather or
not the target belongs to some subset as it is addressed in the $20$-question
framework. These subsets are not constrained to be intervals and the answers
to the queries are noisy. While this situation is well studied for
adaptive querying, this paper is focused on the non adaptive querying
policies based on dyadic questions. The asymptotic minimum achievable distortion under
such policies is derived. Furthermore, a policy named the
\textit{Aurelian\footnote{The policy is named Aurelian to acknowledge
Aurelien Garivier who provided the intuition for it.}}
is exhibited which achieves asymptotically this distortion.

\end{abstract}


\section{Introduction}
\label{sec:intro}

Consider the following problem of localizing a one dimensional deterministic target
$X \in [0;1]$ with a question/answer process. At each step, a subset
of $[0,1]$ is chosen and is questioned: ``whether or not $X$ belongs to this subset". Assume for now that the
answers are truthful, i.e., there is no noise. By repeating this querying, an estimator of $X$ notated $\hat X$ is derived. The performance
of such policy is measured by the supremum distance between the target and
the estimator, $\sup_X |X-\hat X|$. Note that
\begin{equation}
\label{eq:lb}
\sup_X |X-\hat X| \geq \frac{1}{2^{n+1}}
\end{equation}
since at most n bits of information can be learned using n binary questions. Assume now that the subsets are intervals and the policy is non-adaptive. Then,
\begin{equation}
\sup_X |X-\hat X| \geq \frac{1}{2(n+1)}
\end{equation}
which is achieved by choosing intervals of the form $[0,\frac{i}{n+1}]$, for
$1 \leq i \leq n$. Among adaptive policies, the lowest achievable error in (\ref{eq:lb}) is achieved using the
dichotomy policy, consisting in splitting in two intervals of equal size the
current interval containing the target. This performance can also be achieved
using a non adaptive policy with question sets
which are not constrained to be intervals. The dyadic policy, consisting in
querying the bits of $X$ in its expansion in base 2 is an example of a
non-adaptive policy which achieves optimum performance.

Consider now the situation when the answers are noisy according to a memoryless
channel and where $X$ is a sample
from a known distribution. The performance is then measured using the mean square
error. In this case, as shown in \cite{tsiligkaridis2013collaborative},
\begin{equation}
\label{eq:lbn}
E[(X-\hat X_n)^2] \geq A_1\exp(-A_2 n)\,\,\,\,\,\,\,\,\, A_1,A_2>0
\end{equation}
where the expectation is taken over $X$ as well as the noise. $A_1$ and $A_2$ are explicit functions of the entropy of the distribution of $X$ and the characteristics of the noise, respectively. Similarly to the noiseless case, this result is derived by considering the maximum amount of uncertainty reduced in
average by the question/answering process. Now, is this lower bound achievable? 

The bisection
policy consists in choosing the interval $[0,a_n]$ where $a_n$ is the median of the
posterior distribution of $X$ after observing the answer of the $n-1$ first questions.
It is an adaptive policy which achieves the bound in (\ref{eq:lbn}) albeit with different
constants. A weaker statement of (\ref{eq:lbn}) is that the bisection policy achieves
\begin{equation}
\label{eq:limit-n}
\lim_{n \to \infty} \frac{1}{{n}} \ln  E[(X-\hat X_n)^2] = -C
\end{equation}
for some $C>0$. A natural question consists in asking if there exists a non adaptive
policy achieving (\ref{eq:limit-n}) while allowing for querying arbitrary sets,
that is, which are not necessarily intervals. The answer to this question is unknown up to
our knowledge. In order to progress in answering it, we consider a family of
non-adaptive policies. We start with the dyadic policy. Indeed this policy is
optimal in the noiseless case. We also know that the dyadic policy is optimal
for a different loss function which is the differential entropy of the posterior
distribution, see \cite{jedynak2012twenty} and that the differential entropy of
the posterior need to be small for the mean square loss to be small. However,
the reciprocal is not true. We extend the dyadic policy as follows:
In the case where the prior distribution is Uniform, the dyadic questions
correspond to the coefficients of $X$ in its expansion in base 2.
We allow each bit to be queried not only once but an arbitrary  number of times
under the constraint of a fixed total number of questions. This construction is
extended to non Uniform priors by first transforming $X$ into a Uniform random
variable using the transformation $X \mapsto F(X)$ where $F$ is the cumulative
distribution of $X$ and then considering the dyadic questions for $F(X)$. 

\section{Information transmission equivalent problem}
\label{sec:problem_description}

The problem of interest can be mapped to the well-know point to point communication
system \cite{shannon2001mathematical}
which transmits the location of the target through a memoryless channel.
Consider the transmission of a message $x$ in the interval $[0, 1]$
through a binary input, arbitrary output memoryless channel.
The transmission aims at minimizing the mean square error between the input message
and the decoded message. The analysis is presented in the case where the source is
Uniformly distributed. This analysis is then extended to other distributions in
Section~\ref{nonuniform}.
It is assumed that the source provides the binary representation of
the message $x$:
\begin{eqnarray}
x = \sum_{k=1}^{\infty} x_k {2 ^ {-k}}
\label{eq:binary_rep}
\end{eqnarray}
Since the transmission of the infinite bit sequence $x_1, x_2,  ... $, is not
feasible in practice, the message should be quantized by selecting a finite
number of bits, notate by $l$.

\begin{figure}[t!]
  \centering
  \includegraphics[width=7.0cm,height=3.5cm]{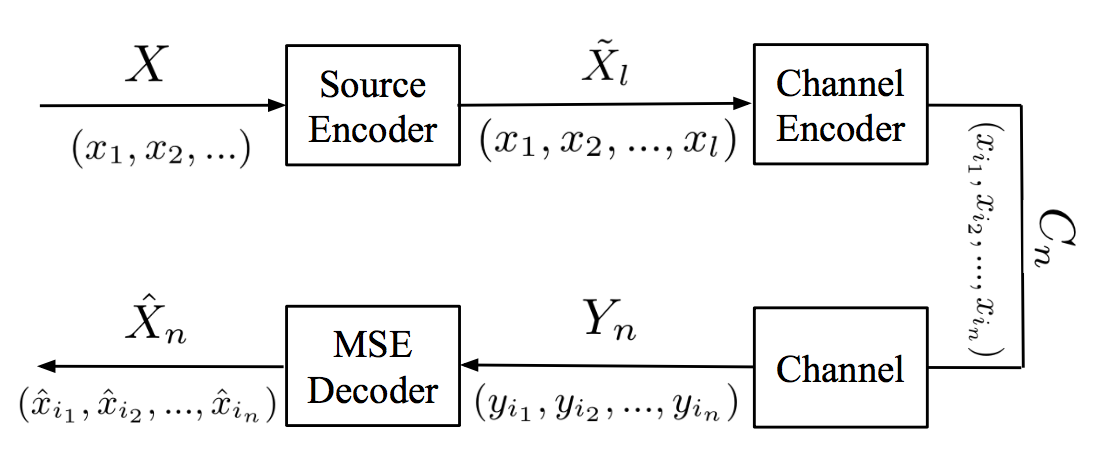}
  \caption{The general transmission interface of a continous message.}
  \label{fig:model}
\end{figure}

The encoding, shown in Figure~\ref{fig:model}, consists in removing the
source redundant information in order to gain compression
efficiency. This
is achieved by a \textit{quantization} scheme. For the uniform source
class, the scalar quantizer where the quantized levels are explicitly
represented via a linear combination of the $x_k$'s has been proved to be
optimal \cite{crimmins1969minimization, mclaughlin1995optimal} for the mean square
error distortion. A source encoder
with fixed rate $l$ is assumed in this paper. The output of the source encoder is
the first $l$ bits of the infinite binary representation of $x$. This quantized
version is denoted as $\tilde{X}_{l}$ in Figure~\ref{fig:model}. The
truncated message is denoted ${\tilde{x}}_{l}=\sum_{k=1}^{l} {x_k {2 ^ {-k}}}$. 

The channel encoder is designed to add redundancy to the
source bit stream to protect the source information from the channel noise.
The bits in the binary representation are of different importance due to the
different weights in the sum \ref{eq:binary_rep}. This motivates the coding
policy which transmits each bit at a different rate. 
\cite{nguyen2012optimal} proposed to transmit each $l$ bits through a single
binary symmetric channel with different transmission rate.
This paper considers the general transmission scheme for which each bit is allowed
to be transmitted a different number of times while the total number of bits to be transmitted
is fixed. 
In Figure~\ref{fig:model}, the
channel encoder sends the $n$ bits codeword
$C_n=(x_{i_1}, x_{i_2}, ..., x_{i_n})$ for which $1 \leq i_j \leq l$.
We denote $q=\max_{j} i_j$ to be the maximum bit index
selected for transmission. Note that $q \leq l$. For any bit $k$, we define
\begin{eqnarray}
t_{n, k} = \sum_{j=1}^{n} \mathbbm{1}_{k}(i_j)
\end{eqnarray}
to be the number of times that bit k is transmitted in the codeword of length $n$.
A transmission policy is then denoted by a vector ${t}_n=(t_{n,1}, t_{n,2}, ...)$
such that $t_{n,k} \geq 0$ for any $k \geq 1$ and $\sum_{k \geq 1} {t_{n,k}} = n$.

The received message $Y_n=(y_{i_1}, y_{i_2}, ..., y_{i_n})$ is different from
the transmitted message due to the channel noise. Here, we use
a \textit{binary input channel} with the following channel posterior
probability:
\begin{eqnarray}
P(Y=y | X=x)=
\begin{cases}
f_1(y) & x=1 \\ f_0(y) & x=0
\end{cases}
\label{eq:channel_posterior}
\end{eqnarray}
where $f_0$ and $f_1$ are point mass functions or densities. 

The decoder is designed to
minimize the end-to-end mean square error (MSE) between the decoded message
${\hat{X}}_n$ and the original message $X$, notated as
$E[( {\hat{X}}_{n} - X)^{2}]$.

The communication system of interest is shown in Figure~\ref{fig:model}. Here,
the source encoder and the channel encoder are two separate entities, but there
is a tight connection between them. This type of \textit{joint source-channel}
coding has been adressed in \cite{goertz2007joint}. This paper considers only
\textit{non-adaptive} transmission policies. The non-adaptive
policies are the policies that uses a one way communication between sender and
receiver. On the other hand the adaptive policies are corresponding to the
system for which there is a
noiseless feedback channel which provides sender the results of the previous
transmission \cite{horstein1963sequential, waeber2013probabilistic}. 


\section{End-To-End Distortion}
\label{sec:end-to-end}

After transmission of $n$ bits, the decoded message ${\hat{X}}_n$ is different
from the source message $X$ due to the quantization and transmission distortion.
The MSE distortion at step $n$ is defined as:
\begin{eqnarray}
E[( {\hat{X}}_{n} - X)^{2}]
= E [ {E[( {\hat{X}}_{n} - X)^{2} | B_n]} ]
\label{eq:distortion}
\end{eqnarray}
here, $E[.]$ denotes statistical expectation and $B_n$ is the history of the
transmissions of the previous bits.


The square error distortion of Eq~\ref{eq:distortion} is minimized when
${\hat{X}}_n = E[X | B_n]$ which lead to the minimum end-to-end distortion:
\begin{eqnarray}
D_n &=& E[(E[X | B_n] - X)^{2} | B_n] \nonumber \\
&=& E[\var(X | B_n)]
\end{eqnarray}
For a given history $B_n$ the transmitted bits $X_i$ and $X_j$ , $i, j \leq n$
and $i \neq j$ are independent and the minimum end-to-end distortion is
therefore, using (\ref{eq:binary_rep}), simplified as:
\begin{eqnarray}
D_n = \sum_{k \geq 1} {2 ^ {-2k}} {E[\var(X_k | B_n)]}
\label{eq:dn}
\end{eqnarray}
where $X_k$ is the random variable corresponding to the $k$'th bit of $X$
in the binary representation of Eq~\ref{eq:binary_rep}. Since each bit $k$
in the above sum is weighted differently, an unequal bit error protection
transmission pattern is desired. The variable $t_{n,k}$ denotes the number
of times that bit $k$ has been transmitted within the first $n$ transmissions.

We now provide bounds on the distortion achieved by any such transmission scheme
or equivalently on any non adaptive policy using the dyadic questions. 
The proof involves classical large deviation inequalities and is provided in the
Appendix. 
\begin{figure}[t!]
  \centering
  \includegraphics[width=8.0cm,keepaspectratio]{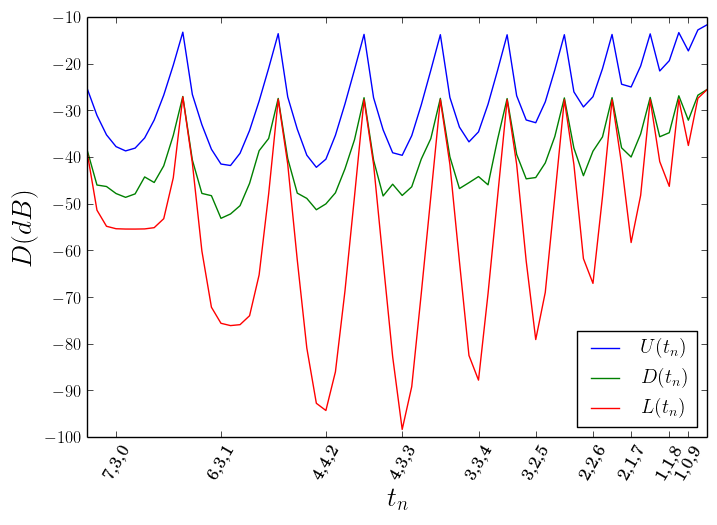}
  \caption{The distortion is bounded from above and below for all transmission policy $\bar{t}_n$
   for $n = 10$ and the maximum depth of $3$.}
  \label{fig:lower_upper}
\end{figure}

\begin{theorem}
\label{th:bounds}
For any transmission pattern $t_n$,
the minimum end-to-end distortion $D_n$ is bounded as follows:
\begin{eqnarray}
L(t_n) \leq D_n \leq U(t_n)
\end{eqnarray}
where,
\begin{eqnarray}
L(t_n) &=& \frac{1}{4} \sum_{k \geq 1} {2 ^ {-2k}} {\exp(-t_{n,k} B(f_1, f_0))} \nonumber \\
U(t_n) &=& \sum_{k \geq 1} {2 ^ {-2k}} {\exp(-t_{n,k} C(f_1, f_0))}
\label{eq:U}
\end{eqnarray}
$C(f_1, f_0)$ is the chernoff information bound \cite{cover2012elements} and
\begin{equation}
B(f_1, f_0) = E\left[\exp\left(-|\ln \frac{f_1}{f_0}(Y)|\right)\right]
\end{equation}
\label{the:theorem_1}
\end{theorem}

Figure~\ref{fig:lower_upper} provides an illustration of theorem \ref{th:bounds}. We show the distortion as well as the lower and
the upper bound for $n=10$ questions and $l=3$ bits source. 
The channel is binary with transitions probabilities $p_{0\to0}=0.9, p_{1\to1}=0.8$.
For this setting, $C=0.77$ and $B=2.08$. There are only $66$ possible
transmissions patterns for which the distortion and the bounds are plotted in
the figure. The experiments are done over $100$ iterations. 
These policies are shown in the x axis. The  policy $t$ which minimizes
$D_{10}(t)$ is $(6,3,1)$ consisting in sending the first bit 6 times,
the second, 3 times and the third once. 

\section{Aurelian Coding Scheme}
\label{sec:non-adaptive}

The optimal non-adaptive policies are the ones which minimize the end-to
-end distortion $D_n$. However, it is difficult to find such optimum transmission
patterns from Eq~\ref{eq:dn}. Instead, we define the \textit{efficient}
policies as follows:

\begin{definition}
An \textit{efficient} policy is a policy 
${t}_n=(t_{n,1}, t_{n,2}, ...)$ that minimizes the upper bound $U$ of the
end-to-end distortion.
\begin{equation*}
 \begin{aligned}
 & \underset{{t}_n}{\text{argmin}}
 & & {\sum_{k \geq 1} { {2 ^ {-2k}} {\exp(-t_{n,k} C(f_1, f_0))}}}\nonumber \\
 & \text{subject to}
 & & \sum_{k=1}^{\infty}{t_{n,k}}=n, t_{n,k} \in \mathbbm{N}\cup \{0\} \\
 &
 \end{aligned}
\end{equation*}
\end{definition}

As Figure~\ref{fig:lower_upper} shows, the upper distortion bound can have many
local minima. Solving such an integer minimization problem might be hard directly.
However,
we can characterize some properties of an \textit{efficient} transmission policy.

\begin{lemma}
For any \textit{efficient} transmission policy $t_n$,
\begin{itemize}
\item{there is no gap between
bit indices selected for transmission, i.e. $t_{n,k} \geq 1$ for $1 \leq k \leq q$,
where $q$ is the last non-zero transmission bit index.}
\item{for any $k_1, k_2 \geq 1$, the difference between transmission values
of the corresponding bits is bounded from above and below:
\begin{eqnarray}
(k_2 - k_1)r - 1 \leq t_{n, k_1} - t_{n, k_2} \leq (k_2 - k_1)r + 1
\end{eqnarray}
where $r = ln(4) / C(f_1, f_0)$.}
\end{itemize}
\label{lab:optimal_properties}
\end{lemma}

These properties are essential to derive the following lower bound on the
minimum end-to-end distortion for all efficient policies.


\begin{theorem}
The logarithm of the minimum end-to-end distortion
of any efficient non-adaptive transmission policy 
goes to $- \infty$  with a rate in $O(-\sqrt{n})$, more specifically:
\begin{eqnarray}
 -A_1 \leq  \lim_{n \to \infty}  \frac{\ln(D_n(t_n))}{\sqrt{n}}  \nonumber
\end{eqnarray}
for each sequence $\{t_n\}_{n \geq 1}$ of efficient policies, where 
\begin{equation*}
A_1=\min(\sqrt{2}(\frac{\ln(4)}{C(f_1,f_0)} + 1)B(f_1,f_0), \ln(4))
\end{equation*}
\label{the:efficient_bound}
\end{theorem}

Even though we derived some \textit{necessary} important properties for policies
that achieve the infimum of the upper bound, we did not provide an explicit
efficient non-adaptive transmission policy. Our strategy is the following:
We propose a policy, called the \textit{Aurelian} policy, derived from
the minimization of (\ref{eq:U}) for sequences of real numbers $t_n$ (Of course we then
adapt the $t$'s such that we obtain a sequence of integers). 

\begin{definition}
\textbf{Aurelian Policy}:
\begin{eqnarray}
q &=&  \floor*{\sqrt{\frac{2n}{r} + \frac{1}{4}} -\frac{1}{2}} \nonumber \\
t_{n, 1} &=& qr \nonumber \\
t_{n, k} &=& t_{n,1} - (k - 1)r \nonumber \\
r &=& \floor*{\ln(4) / C(f_1, f_0)}
\end{eqnarray}
\end{definition}


At this stage, even if the choice of the \textit{Aurelian} policy is based on
some "good" intuitive reasons ( the minimization of the continuous problem),
we do not have any insurance on the rate of convergence to $0$ of the distortion
rate when this policy is used, compared to the rate of convergence of an efficient
non-adaptive transmission policy. Theorem 2 address this
this problem. First of all, we will derive a upper bound for the convergence rate
to $-\infty$ of the  distortion rate of the \textit{Aurelian} policy. Having this
rate, we will show that it is comparable to the  distortion rate of an efficient
non-adaptive transmission policy. In other words, we can conclude that we don't
loose "too much" by following an \textit{Aurelian} policy instead of deriving an
explicit non-adaptive transmission policy.

\begin{theorem}
The logarithmic rate of convergence of the \textit{Aurelian} policy is no more
than $O(-\sqrt{n})$. More precisely,
\begin{eqnarray}
\limsup_{n\to\infty} \frac{\ln(D_n)}{\sqrt{n}}  \leq - A_2
\end{eqnarray}
where $A_2 = \sqrt{2r} C(f_1, f_0)$.
\label{the:aurelian}
\end{theorem}

And finally,

\begin{corollary}
If we denote by $D_n$ the distortion rate of the \textit{Aurelian} policy, and $D_n^*$ the distortion rate of any efficient non-adaptive transmission policy, then we have:
\begin{eqnarray}
\liminf_{n \to \infty}\frac{\ln(D_n)}{\ln(D_n^*)} \geq \frac{A_1}{A_2} > 0
\end{eqnarray}
for $A_1$ and $A_2$ given in Theorem~\ref{the:efficient_bound} and
Theorem~\ref{the:aurelian}, respectively.
\end{corollary}

\begin{figure}[t!]
  \centering
  \includegraphics[width=8.0cm,keepaspectratio]{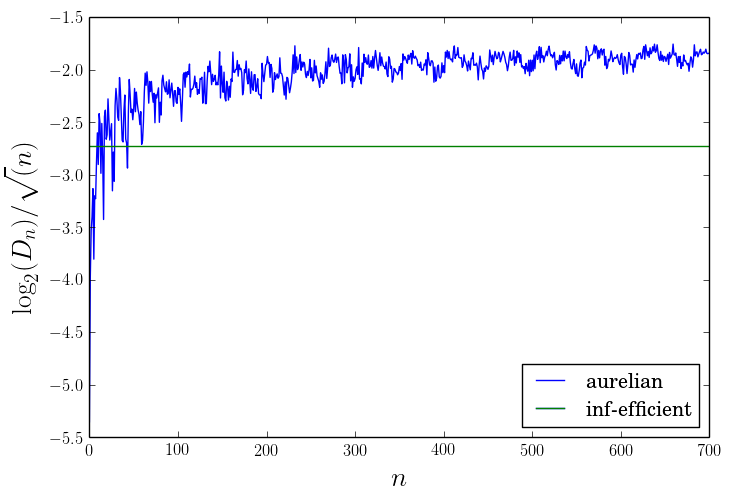}
  \caption{The normalized distortion versus the number of transmissions as well as the lower bound, i.e the constant $-A_1$ provided in Theorem 2. }
  \label{fig:asymptotic}
\end{figure}

The blue curve in Figure~\ref{fig:asymptotic} shows the asymptotic behavior
of the aurelian policy. As this figure shows, after about $300$ transmissions,
the receiver is able to decode the message with very high confidence. It is also
worth to mention that the rate of convergence is of order $1 / \sqrt{n}$.

\section{Non-Uniform Distribution}
\label{nonuniform}

Note that the lower bound of Theorem~\ref{the:efficient_bound} generalizes to random variables
for which the cumulative distribution is  Lipschitz continous. Indeed in this case, 
by definition, there exists a constant $k>0$ such that 
\begin{eqnarray}
(F(u_1) - F(u_2))^{2} \leq k ({u_1} - u_2)^2
\end{eqnarray}
The cumulative disttribution of any random varaible is a uniform random variable in
interval $[0,1]$. Instead of locating target $X$, we can search for the target $F(X)$.
Lets $F_n$ be the estimated location after $n$ steps and let $\hat{X}_n = F^{-1}(F_n)$.
\begin{eqnarray}
(F_n - F(X))^{2} \leq k (\hat{X}_n - X)^2 \nonumber
\end{eqnarray}
The lower bound of the distortion is then derived using
Theorem~\ref{the:efficient_bound}.

\section{Conclusion}
\label{conclusion}

The problem of noisy target localization under mean square distortion is addressed through an iterative binary question/answering process. Among non-adaptive policies, the policies involving dyadic questions were considered. It was shown that the logarithm of the mean square distortion is $O(-\sqrt{n})$ for $n$ large enough. An explicit policy achieving this rate was exhibited. We finally conjecture that this policy is optimal among all non-adaptive policies.

\bibliographystyle{IEEEbib}
\bibliography{refs}

\section{Appendix}
\label{appendix}
\subsection{Proof of Theorem 1:}
\subsubsection{Upper bound}
\textit{Proof:}
We first prove the follwoing inequality:
\begin{eqnarray}
E[var(X_k | B_n)] &\leq& \exp(-t_{n,k} C(f_1, f_0))
\end{eqnarray}
the upper bound is then derived by considering the fact that for any random
variable $A$ and scalar $b$, if $A \leq b$, then $E[A] \leq b$. Note:
\begin{eqnarray}
Var[X_k | B_n] &=& P(X_k = 1 | B_n) (1 - P(X_k = 1 | B_n)) \nonumber \\
&\leq& min(P(X_k=0 | B_n), P(X_k =1 | B_n)) \nonumber \\
&=&P({\tilde{X}}_k \neq X_k | B_n) \nonumber
\end{eqnarray}
where ${\tilde{X}}_k$ is the mode of $p(X_k|B_n)$. The above equation can be
simplified further as:
\begin{eqnarray}
P({\tilde{X}}_k \neq X_k | B_n) &=& P({\tilde{x}}_k=1 | B_n, X_k=0) P(X_k=0) \nonumber \\
&+&P({\tilde{X}}_k=0|B_n,X_k=1)P(X_k=1) \nonumber
\end{eqnarray}
where,
\begin{eqnarray}
P({\tilde{X}}_k=1 | B_n, X_k=0)&=& P(\sum_{j=1}^{n}{\mathbbm{1}_{k}(i_j) \log(\frac{f_1}{f_0}(y_{i_j}) > 0)}) \nonumber \\
&=& P(\exp(s \sum_{j=1}^{n} \mathbbm{1}_k(i_j) \log(\frac{f_1}{f_0}(y_{i_j}))) > 1)\nonumber \\
&\leq& E[\exp(s \sum_{j=1}^{n} \mathbbm{1}_{k}(i_j) \log(\frac{f_1}{f_0}(y_{i_j})))] \nonumber \\
&=& \exp(-t_{n,k} C(f_1, f_0)) \nonumber
\end{eqnarray}
here, $\mathbbm{1}_k(i_j)$ is the indicator of sending bit $k$ at transmission step
$i_j$. The
second equality is hold for any positive value $s$. The third inequality is derived
using Markov inequality and the final equality holds based on the fact that
$\sum_{j=1}^{n} {\mathbbm{1}_{k}(i_j)} = t_{n,k}$ by defnition.
The fact that each transmission is independent of the rest and have same i.i.d
distribution is also used for deriving the expectation of the last inequality.

\subsubsection{Lower bound}

Next, Lets define the posterior function $p_{n,k} = P(X_k=1|B_{n})$.
\begin{lemma}
The posterior denisty on bit $k$ at step $n$ is $p_{n,k} = p_{n-1, k}$
whenever $\mathbbm{1}_{k}(i_n)=0$. Otherwise the posterior is a function
of the posterior at previous transmission step:
\begin{eqnarray}
p_{n,k} =
\frac{f_1(y_{i_n})p_{n-1,k}}{f_1(y_{i_n}) p_{n - 1, k} + f_0(y_{i_n}) (1 - p_{n - 1, k})}
\end{eqnarray}
\label{lem:post_lemma}
\end{lemma}
\textit{Prrof:} First note that since bits are independent random variables,
sending a bit different from $k$ will not change the posterior at step $n$,
thus $p_{n,k} = p_{n - 1, k}$ when $i_n \neq k$. In the other hand, whenever
the bit $k$ was sent through the channel at the $n$'th transmission:
\begin{eqnarray}
p_{n, k} &=& P(X_k = 1 | B_n) \nonumber \\
&=& \frac{P(y_{i_n} | X_k=1, B_{n-1}) P(X_k=1 | B_{n-1})} {P(y_{i_n} | B_{n - 1})} \nonumber
\end{eqnarray}
where,
\begin{eqnarray}
P(y_{i_n} | B_{n - 1}) &=& \sum_{u \in \{0,1\}} P(y_{i_n} , X_k = u | B_{n-1}) \nonumber \\
&=& P(y_{i_n} | X_k=1, B_{n-1}) P(X_k=1 | B_{n-1}) \nonumber \\
&+& P(y_{i_n} | X_k=0, B_{n-1}) P(X_k=0 | B_{n-1}) \nonumber \\
&=& f_1(y_{i_n}) p_{n - 1, k} + f_0(y_{i_n}) (1 - p_{n - 1, k})
\label{eq:p_n_q}
\end{eqnarray}
\begin{lemma}
If $i_n = k$
\begin{eqnarray}
p_{n,k}(1 - p_{n,k}) \geq p_{n-1, k}(1 - p_{n-1, k}) e ^ {-\mathbbm{1}_{k}(i_n) |\log(\frac{f_1}{f_0}(y_{i_n})|)}
\end{eqnarray}
\label{lem:lemma_2}
\end{lemma}
\textit{Proof:} For $i_n=k$.
\begin{eqnarray}
p_{n,k}(1-p_{n,k}) &=& \frac{f_1(y_{i_n})f_0(y_{i_n})p_{n-1,k}(1-p_{n-1,k})}{[f_1(y_{i_n}) p_{n - 1, k} + f_0(y_{i_n}) (1 - p_{n - 1, k})]^2} \nonumber \\
&\geq& \frac{f_1(y_{i_n})f_0(y_{i_n})p_{n-1,k}(1-p_{n-1,k})}{f_1(y_{i_n}) \mathbbm{1}_{\frac{f_1(y_{i_n})}{f_0(y_{i_n})} \geq 1} + f_0(y_{i_n}) \mathbbm{1}_{\frac{f_0(y_{i_n})}{f_1(y_{i_n})} \geq 1}
}\nonumber \\
&=& \frac{p_{n-1,k}(1-p_{n-1,k})}{\frac{f_0(y_{i_n})}{f_1(y_{i_n})} \mathbbm{1}_{\frac{f_1(y_{i_n})}{f_0(y_{i_n})} \geq 1} + \frac{f_1(y_{i_n})}{f_0(y_{i_n})} \mathbbm{1}_{\frac{f_0(y_{i_n})}{f_1(y_{i_n})} \geq 1}} \nonumber \\
&=& p_{n-1, k}(1 - p_{n-1, k}) \exp(- |\log \frac{f_1}{f_0}(y_{i_n})|)\nonumber
\end{eqnarray}

\begin{corollary}
After transmission of $n$ bits,
\begin{eqnarray}
p_{n,k}(1-p_{n,k}) \geq p_{0,k}(1 - p_{0,k}) \exp(-t_{n,k} B(f_1, f_0))
\end{eqnarray}
where, $B(f_1, f_0) = E[|\log (\frac{f_1(y)}{f_0(y)})|]$.
\label{cor:cor_1}
\end{corollary}

\textit{Proof:}
Using Lemma~\ref{lem:lemma_2} recursively:
\begin{eqnarray}
p_{n,k}(1 - p_{n,k})
&\geq& p_{0,k}(1 - p_{0,k}) \prod_{j=1}^{n} \exp({-\mathbbm{1}_{k}(i_j) |\log(\frac{f_1(y_{i_j})}{f_0(y_{i_j})})|}) \nonumber \\
&=& p_{0,k}(1 - p_{0,k}) \exp(- \sum_{j=1}^{n} {\mathbbm{1}_{k}(i_j) |\log(\frac{f_1(y_{i_j})}{f_0(y_{i_j})})|}) \nonumber \\ \nonumber
\end{eqnarray}
Taking expectation from both side and using Jensen inequality:
\begin{eqnarray}
E[var(X_k | B_n)] &=&E[p_{n,k}(1 - p_{n,k})] \nonumber \\
&\geq& p_{0,k}(1 - p_{0,k}) \exp(- t_{n,k} { E[|\log(\frac{f_1(y)}{f_0(y)})|]}) \nonumber \\ \nonumber
\end{eqnarray}
where the last equation derived based on the fact that there is only $t_{n,k}$ steps
that bit $k$ has been sent and these steps are i.i.d.

\subsection{Optimal Properties}

\textbf{Property 1}:
Lets assume there is a bit $k_1$ such that $t_{n,k_1}=0$ and let $k_2$ be
the first index greater $k_1$ such that $t_{n, k_2} > 0$. Lets $t'_{n}$ be
a new policy made from the earlier policy $t_n$, such that
\begin{eqnarray}
t'_{n, k}=
\begin{cases}
t_{n, k_2}\,\,\,& k=k_1 \\ 0\,\,\, & k=k_2 \\ t_{n,k} & k\not \in \{k_1, k_2\}
\end{cases}
\end{eqnarray}
we claim $U(t'_n) \leq U(t_n)$:
\begin{eqnarray}
U({t}'_n) - U({t}_n) &=& \sum_{k \geq 1} {{2 ^ {-2k}} (\exp(t'_{n,k}C) - \exp(t_{n,k}C))} \nonumber \\
&=& {2 ^ {-2 k_1}}(e ^{-t_{n, k_2}C} - 1)+ {2 ^ {-2 k_2}}(1 -{e ^ {-t_{n, k_2}C}}) \nonumber \\
&=&(e ^{-t_{n, k_2}C} - 1)({2 ^ {-2 k_1}} - {2 ^ {-2 k_2}}) \nonumber \\
&\leq& 0 \nonumber
\end{eqnarray}
Since $0 \leq C \leq 1$ and $k_1 < k_2$.n

Let us call the transmission policy $t'_n$ derivable from $t_n$ by a
$(k_1, k_2)$-move when 
\begin{eqnarray}
t'_{k,n} =
\begin{cases}
t_{k, n} + 1 & \text{if}\,\,\, k=k_1 \\ t_{k, n} - 1 & \text{if}\,\,\, k=k_2 \\ t_{k, n}  & \text{Otherwise}
\end{cases}
\nonumber
\end{eqnarray}
here it is assumed that $t_{n,k_2} \geq 1$, otherwise the above move can not be
defined.
An optimal transmission policy is such that no further $(k_1, k_2)$-move
can lead to a transmission policy with lower upper bound distortion. This is the
key idea to prove the following lemma:

\noindent \textbf{Property 2}:
Lets $t_n$ be a coding policy and $t'_n$ be the policy
derived by a $(k_1, k_2)$-move between two bit indices $k_1$ and $k_2$
such that $t_{n, k_2} \geq 1$.
This move reduce the upper bound values if
$U(t_n) \geq  U(t'_n)$, simplification of both side
prove the following lower bound inequality
\begin{eqnarray}
(k_2 - k_1)r - 1 \leq t_{n, k_1} - t_{n, k_2} \nonumber
\end{eqnarray}
Similarly, the other bound can be derived by considering situations that a
$(k_2, k_1)$-move can reduce the upper bound.

\begin{corollary}
Using the above properties, we can derive the following bounds for the last
transmitted bit index $q$ and the transmission number of first bit, $t_{n,1}$:
\begin{eqnarray}
t_{n,1} &\leq& q(r + 1) \nonumber \\
q &\leq& \sqrt{2n + 1/2} - 1/4 \nonumber
\end{eqnarray}
\label{cor:properties_corol}
\end{corollary}
The proof is very straigtforward using the properties and note
$\sum_{k=1}^{q} t_{n,k}=n$.

\subsection{Asymptotic Behavior of Infimum of the Efficient Policies}
Lets $t_n$ be an efficient policy. For such policy
\begin{eqnarray}
L(t_n) &=& \frac{1}{4} [ \sum_{k \geq 1} {2 ^ {-2k} \exp(-t_{n,k}B)}] \nonumber \\
&=&\frac{1}{4} [\sum_{k=1}^{q} {\exp(\underbrace{(-k \log(4) - t_{n,k} B)}_{v_k})} + \sum_{k>q} {2^{-2k}} ]\nonumber \\
&=&\frac{1}{4} [{e ^ {v_1}} \sum_{k=1}^{q} {\exp({v_k - v_1})} + {2 ^ {-2 (q + 1)}}\frac{4}{3} ] \nonumber \\
&\geq& \frac{1}{4} [e ^ {v_1 - B} \frac{e ^ {-q \alpha} - 1}{e^{- \alpha}} + \frac{1}{3} e ^ {-q log(4)}]
\end{eqnarray}
where $\alpha = log(4)(1 + \frac{B}{C}) > log(4)$, since $B, C \geq 0$. Using the
Corollary~\ref{cor:properties_corol}:
\begin{eqnarray}
v_1 \geq  \log(4) - q(\frac{\log(4)}{C} + 1)B
\end{eqnarray}
let $n \to \infty$ and using the upper bound of $q$,
\begin{eqnarray}
\lim_{n \to \infty} \inf_{t_n \in \mathbbm{T}^{*}} \frac{\log(L(t_n))}{\sqrt{n}} \geq -A_1
\end{eqnarray}
where $A_1 = \min(\sqrt{2}(\frac{\log(4)}{C} + 1)B, \log(4))$ which complete the proof.

\subsection{Asymptotic Behavior of Supremum of the Aurelian Policies}
For an Aurelian policy $t_n$:
Lets $t_n$ be an efficient policy. For such policy
\begin{eqnarray}
U(t_n) &=& \sum_{k \geq 1} {2 ^ {-2k} \exp(-t_{n,k}C)} \nonumber \\
&=&\sum_{k=1}^{q} {\exp({(-k \log(4) - t_{n,k} C)})} + \sum_{k>q} {2^{-2k}} \nonumber \\
&=&\sum_{k=1}^{q} {\exp({- \log(4) - t_{n,1} C})} + {2 ^ {-2 (q + 1)}}\frac{4}{3}\nonumber \\
&=& q \exp(- \log(4) - q r C) + \frac{1}{3} \exp(-q \log(4))
\end{eqnarray}
For $n \to \infty$, since $r C \leq \log(4)$:
\begin{eqnarray}
\lim_{n \to \infty} \sup_{t_n \in \mathbbm{T}^{*}} \frac{\log(U(t_n))}{\sqrt{n}} \leq -A_2
\end{eqnarray}
where $A_2 = \sqrt(2r) C$.

\end{document}